\documentclass[twocolumn]{aastex63}

\def\lapp{\ifmmode\stackrel{<}{_{\sim}}\else$\stackrel{<}{_{\sim}}$\fi}
\def\gapp{\ifmmode\stackrel{>}{_{\sim}}\else$\stackrel{>}{_{\sim}}$\fi}
\usepackage{multirow}
\usepackage{color}
\usepackage{amsmath}
\usepackage{soul}
\usepackage{hyperref}
\urldef{\footurl}\url{https://www.cosmos.esa.int/web/xmm-newton/sas-thread-epic-filterbackground}
\shorttitle{}
\shortauthors{}
\begin{document}

\title{X-ray constraints on the spectral energy distribution of the $z=5.18$ blazar SDSS~J013127.34$-$032100.1}

\correspondingauthor{Hongjun An}
\email{hjan@cbnu.ac.kr}

\author{Hongjun An}
\affiliation{Department of Astronomy and Space Science, Chungbuk National University, Cheongju, 28644, Republic of Korea}
\author{Roger W. Romani}
\affiliation{Department of Physics/KIPAC, Stanford University, Stanford, CA 94305-4060, USA}

\begin{abstract}
We report on X-ray measurements constraining the spectral energy distribution (SED) of the high-redshift $z=5.18$ blazar SDSS~J013127.34$-$032100.1 with new {\it XMM-Newton} and {\it NuSTAR} exposures. The blazar's X-ray spectrum is well fit by a power law with $\Gamma=1.9$ and $N_{\rm H}=1.1\times10^{21}\rm \ cm^{-2}$, or a broken power law with $\Gamma_l=0.5$, $\Gamma_h=1.8$, and a break energy $E_b=0.7$\,keV for an expected absorbing column density of $N_{\rm H}=3.6\times 10^{20}\rm \ cm^{-2}$, supported by spectral fitting of a nearby bright source. No additional spectral break is found at higher X-ray energies (1--30\,keV). We supplement the X-ray data with lower-energy radio-to-optical measurements and {\it Fermi}-LAT gamma-ray upper limits, construct broadband SEDs of the source, and model the SEDs using a synchro-Compton scenario. This modeling constrains the bulk Doppler factor of the jets to $\ge$7 and $\ge$6 (90\%) for the low- and high-$N_{\rm H}$ SEDs, respectively. The corresponding beaming implies $\ge$130 (low $N_{\rm H}$) or $\ge$100 (high $N_{\rm H}$) high-spin supermassive black holes similar to J0131 exist at similar redshifts.

\end{abstract}


\section{Introduction}
\label{sec:intro}
Supermassive black holes exist even at high redshifts \citep[e.g., $z\ge5$;][]{fnls+01}. Rapidly spinning black holes may power bipolar jets via the Blandford-Znajek mechanism \citep[][]{bz77}; when these jets lie close to the Earth line of sight (LoS) these are visible as bright `blazars' \citep[][]{up95}. These jets are relativistic, accelerating high energy particles which interact with magnetic field and ambient photons, and producing broadband emission which is further enhanced by Doppler beaming due to relativistic bulk motion. This emission can be bright across the electromagnetic spectrum, allowing detection to very high redshifts \citep[e.g., $z=5.48$;][]{rsgp04}.

Blazars' broadband spectral energy distributions (SEDs) exhibit a characteristic double-hump spectrum, one at low frequencies ($<$X-rays) and another at higher frequencies. The former is believed to be produced by synchrotron radiation
of the accelerated particles, and
the latter by inverse-Compton upscattering of soft photons from the synchrotron radiation (synchrotron self-Compton; SSC),
and the torus, disk, and broad line region (external Compton; EC) by the jet particles \citep[e.g.,][]{bms97}.
This scenario has been used to model SEDs of high-$z$ blazars \citep[][]{r06,stgp+13,gtsg15,ar18}.
The emission is relativistically beamed due to the bulk motion of the jet
by the Doppler factor $\delta=1/[\Gamma_D(1-\sqrt{1-1/\Gamma_D^2}\mathrm{cos}\theta_V)]$,
where $\Gamma_D$ is the jet Lorentz factor, and $\theta_V$ is the viewing angle;
these parameters are crucial for estimating the jet luminosity and beaming factor,
and are determined from the SED peak frequencies and amplitudes via SED modeling.
The shape of the SED is determined by conditions in the emission zone, is sensitive to variations in these conditions, and
is modified by the environment (e.g., soft photon fields) along the LoS.
Therefore the SEDs probe the source emission zone properties as well as the extragalactic background via emission \citep[e.g.,][]{gtsg15}, variability \citep[e.g.,][]{lrfk+18} and absorption \citep[e.g.,][]{fermiebl} studies.

\begin{figure*}
\centering
\includegraphics[width=170mm]{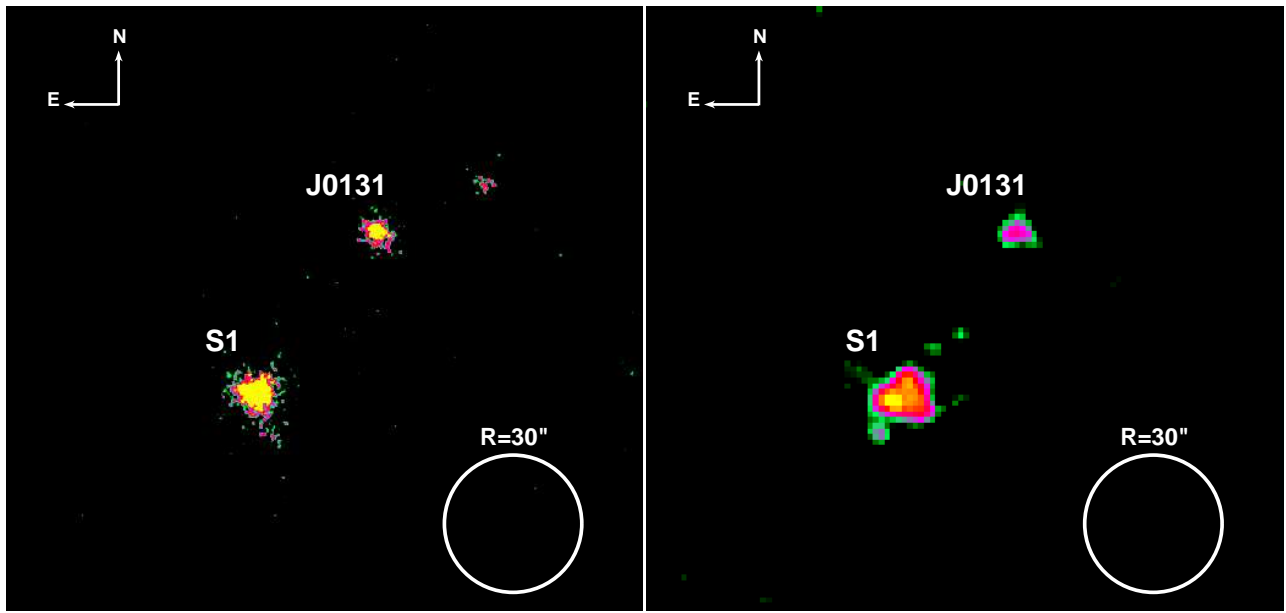} \\
\figcaption{0.2--10\,keV and 3--30\,keV images taken with {\it XMM}/MOS (left) and {\it NuSTAR} (right).
J0131 is detected up to $\sim$30\,keV, and another brighter source (S1) is visible at $\sim$90$''$ south east of J0131. The images are smoothed and scales are adjusted for better legibility. An $R=30''$ white circle is drawn in the lower right corner for reference. Note that the south source ``S1'' appears to have an extended emission (south-west jet-like structure) in the {\it XMM} image. Our investigations with the {\it XMM}/OM images suggest that the structure may be produced by contamination from another faint point source near the tip of the structure.
\label{fig:fig1}}
\end{figure*}

High-redshift (high-$z$) blazars with well-measured SEDs are very useful in understanding the early universe and its evolution. In particular, SED studies of high-$z$ blazars provide estimates of their jet $\delta$'s \citep[e.g.,][]{r06,gtsg15,ar18}. For these high-power sources the characteristic double-hump spectrum has a synchrotron peak in the far-IR to optical band and
a high-energy Compton peak in the X-ray to $\gamma$-ray band. The X-rays are thought to probe the SSC SED
and possibly the rising tail of the EC emission, and the peak locations of the high-energy humps with respect to the
low-energy synchrotron one are sensitively dependent on $\Gamma_D$. Thus,
good coverage of these peaks is required for $\Gamma_D$ measurements. Since the black hole masses can be estimated from SED and spectroscopic studies and since a strong jet is generally assumed to be related to high spin, the population of luminous high-$z$ blazars tells us much about the origin and growth mechanisms \citep[e.g.,][]{bv08} of high mass, high spin holes in the early universe.
While such blazars are rare (only four at $z>5$ have been reported to date), with a beaming correction $\approx 2\Gamma_D^2$, they represent a substantial population.

SDSS~J013127.34$-$032100.1 (J0131 hereafter) is an optically luminous high-$z$ blazar \citep[$z=5.18$;][]{ywwy+14}. Its broadband SED exhibits the characteristic double-hump structure with strong disk emission in the optical band. Previous SED modelings, using existing {\it Neil-Gehrels-Swift} X-ray observatory \citep[XRT;][]{bhnk+05} data, have been used to infer a modest jet Doppler factor \citep[$\delta\ge10$;][]{gtsg15}. These X-ray data only poorly constrained the Compton component, so Doppler factor constraints were quite weak. We therefore observed the source with {\it XMM-Newton} \citep[][]{jlac+01} and {\it NuSTAR} \citep[][]{hcc+13} to improve the X-ray SED characterization. We then combine multi-waveband data, construct broadband SEDs, and model with a synchro-Compton code \citep[][]{bms97} to better constrain the jet Doppler factor.

We describe the data reduction and analyses in \S\ref{sec:sec2}. The broadband SEDs are constructed and modeled in \S\ref{sec:sec3}. We conclude by discussing the population implications in \S\ref{sec:conclusion}.

\section{Data reduction and Analysis\label{sec:sec2}}

\subsection{X-ray data reduction\label{sec:sec2_1}}
We obtained contemporaneous 50\,ks {\it XMM} and 100\,ks {\it NuSTAR} exposures on MJD~58847.3. The {\it XMM} data are processed with the {\tt emproc} and {\tt epproc} tools of SAS~20190531\_1155, and particle flares are removed following the standard procedure.\footnote{{https://www.cosmos.esa.int/web/xmm-newton/sas-thread-epic\\-filterbackground}}
The exposures after this process are 43\,ks, 43\,ks and 32\,ks for the MOS1, MOS2 and PN data, respectively. We process the {\it NuSTAR} data using the {\tt nupipeline} tool in HEASOFT~6.26.1, setting {\tt saamode=strict} and {\tt tentacle=yes} as recommended by the {\it NuSTAR} SOC.\footnote{http://www.srl.caltech.edu/NuSTAR\_Public/NuSTAROpera\\tionSite/SAA\_Filtering/SAA\_Filter.php}
This reduces the exposure to 80\,ks. J0131 is faint but clearly visible in the cleaned images, and a brighter second source is detected $\sim90''$ south east of J0131 (Fig.~\ref{fig:fig1}). We note that the south Atlantic anomaly (SAA) filter setting for the {\it NuSTAR} process does not have large impact on our results below.

\begin{table*}[t!]
\caption{Summary of X-ray spectral fits}
\label{ta:ta1}
\begin{center}
\scriptsize{
\begin{tabular}{lccccccccc}
\hline
Data$^{\rm a}$  & model &  Energy & $N_{\rm H}$      & $\Gamma_l$  & $E_{\rm b}$  & $\Gamma_h$ & $F_{\rm 3-10\rm \ keV}$ & $\chi^2$/dof & Comments \\ 
      &  &  (keV)  & ($10^{22}\rm cm^{-2}$)  &      & (keV)    &     & ($10^{-14}\rm \ erg\ cm^{-2}\ s^{-1}$) & \\ \hline
X     & PL & 0.2--10  & $0.11\pm0.02$ & $1.89\pm0.07$ & $\cdots$ & $\cdots$ & $6.1\pm0.6$ & 69/90 & large $N_{\rm H}$ \\
X     & BPL & 0.2--10  & $0.17\pm0.05$ & $2.3\pm 0.4$ & $1.4\pm 0.4$ & $1.8\pm 0.1$ & $6.6\pm 0.7$ & 67/88 & large $N_{\rm H}$ \\
X     & PL & 0.2--10  & 0.036$^{\rm b}$   & $1.63\pm0.04$  & $\cdots$   & $\cdots$ & $8.0\pm0.6$ & 98/91 & small $N_{\rm H}$ \\
X     & BPL & 0.2--10  & 0.036$^{\rm b}$   & $0.5\pm 0.4$  & $0.7\pm 0.1$   & $1.78\pm 0.05$ & $6.7\pm 0.6$ & 70/89 & small $N_{\rm H}$ \\
N     & PL & 3--30    & 0.11$^{\rm b}$    & $1.7\pm 0.3$  & $\cdots$ & $\cdots$ & $7.1\pm1.4$ & 23/27 & insensitive to $N_{\rm H}$ \\ \hline
X,S,N & PL & 0.2--30  & $0.11\pm0.02$ & $1.86\pm 0.07$ & $\cdots$ & $\cdots$ & $6.1\pm0.6$ & 337/382 & large $N_{\rm H}$ \\
X,S,N & BPL & 0.2--30  & 0.036$^{\rm b}$   & $0.7\pm0.3$  & $0.7\pm 0.1$ & $1.78\pm 0.05$ & $6.4\pm0.6$ & 337/381 & small $N_{\rm H}$ \\ \hline
\end{tabular}}
\end{center}
\footnotesize{
$^{\rm a}$ X: {\it XMM}, N: {\it NuSTAR}, S: {\it Swift}\\
$^{\rm b}$ Frozen}
\end{table*}

\subsection{Spectral analysis for the {\it XMM} data\label{sec:sec2_2}}
We first measure 0.2--10\,keV spectra of J0131 using the {\it XMM} data. We extract events within $R=16''$ and $R=32''$ radius circles for the source and background, respectively, and compute the corresponding response files with the {\tt arfgen} and {\tt rmfgen} tasks of SAS. The source spectra are grouped to have at least 20 counts per bin, and we fit the spectra with an absorbed power-law (PL) model using {\tt wilm} abundance \citep[][]{wam00} and {\tt vern} cross section \citep[][]{vfky96} in {\tt XSPEC}~12.10.1. Because {\it XMM} provides high-quality data at low energies ($<$1\,keV), we allow the absorbing column density $N_{\rm H}$ to vary and then cross-check the inferred value with Galactic HI and optical extinction measurements.

We find that the model fits the data well ($\chi^2$/dof=69/90) with $N_{\rm H}=(1.1\pm0.2)\times 10^{21}\rm \ cm^{-2}$, $\Gamma = 1.89\pm 0.07$ and unabsorbed flux $F_{3-10\rm \ keV}=(6.1\pm0.6)\times 10^{-14}\rm \ erg\ cm^{-2}\ s^{-1}$ (Table~\ref{ta:ta1}). We also try to fit the spectra with a broken power-law (BPL) model, but the model does not improve the PL fit at all (Table~\ref{ta:ta1}), meaning that a simple PL is adequate. While this model is quite acceptable, we find that the fit $N_{\rm H}$ value is significantly larger than the $\sim$$3.5\times 10^{20}\rm \ cm^{-2}$ expected in this area from radio HI mapping.\footnote{\url{https://heasarc.gsfc.nasa.gov/cgi-bin/Tools/w3nh/w3nh.pl}} The finer scale of the PannSTARS2 dust maps \citep[][]{gszs+19} shows patchy extinction in this direction, with low absorption at the blazar position, but values as large as $N_{\rm H}=10^{21}\rm \ cm^{-2}$ within 30$^\prime$.

\begin{figure}
\centering
\includegraphics[width=85mm]{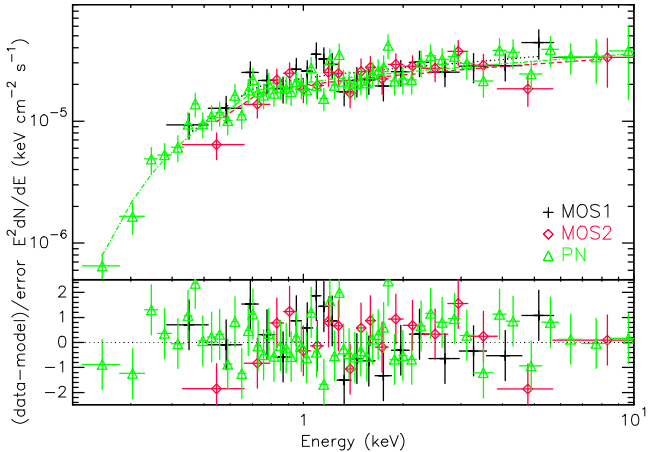} \\
\figcaption{0.2--10\,keV SED measured with {\it XMM} MOS1 (black), MOS2 (red), and PN (green). {\it Top}: SED data (data points) and the best-fit BPL model (lines). {\it Bottom}: fit residuals.
\label{fig:fig2}}
\end{figure}

A check on the $N_{\rm H}$ in this direction is provided by a nearby bright(er) source $\sim$90$''$ south east of J0131 (``S1'' in Fig.~\ref{fig:fig1}). We extract the source and background spectra with $R=16''$ and $R=32''$ radius circles, and generate the corresponding response files using the procedure described above. The observed X-ray spectra of this source appear to be a PL or a BPL, so we fit the spectra with both models. The PL fit gives $N_{\rm H}=(2.3\pm0.4)\times 10^{20}\rm \ cm^{-2}$, $\Gamma=1.80\pm 0.03$, and $F_{3-10\rm\ keV}=(1.36\pm0.07)\times 10^{-13}\rm \ erg\ cm^{-2}\ s^{-1}$ ($\chi^2$/dof=182/220), while the BPL model has $N_{\rm H}=(3.6\pm0.6)\times 10^{20}\rm \ cm^{-2}$,
$\Gamma_{l}=1.98\pm 0.07$, $\Gamma_{h}=1.59\pm0.08$, $E_{\rm b}=1.9\pm0.4$\,keV, and $F_{3-10\rm\ keV}=(1.5\pm0.1)\times 10^{-13}\rm \ erg\ cm^{-2}\ s^{-1}$ ($\chi^2$/dof=168/218). The BPL provides a significantly better fit ($f$-test probability of $2\times 10^{-4}$). Note that the $N_{\rm H}$ values inferred from these models (toward S1) are significantly lower than that inferred for J0131, and  the BPL S1 model $N_{\rm H}$ agrees very well with the Galactic HI estimate. Adding {\it NuSTAR} data in the fits does not alter the results for S1.

We therefore have fit the {\it XMM} J0131 spectra with $N_{\rm H}$ fixed at $3.6\times 10^{20}\rm \ cm^{-2}$. A PL model with $\Gamma=1.63\pm0.04$ has $\chi^2$/dof=98/91 but residual trends at low and high energies are clearly visible. Accordingly, we then apply a BPL model and find that a model with $\Gamma_l=0.5\pm0.4$, $\Gamma_h=1.78\pm0.05$, $E_{\rm b}=0.7\pm0.1$\,keV, and $F_{\rm 3-10\rm \ keV}=(6.7\pm0.6)\times 10^{-14}\rm \ erg\ cm^{-2}\ s^{-1}$ (Table~\ref{ta:ta1} and Fig.~\ref{fig:fig2}) provides a significantly better fit ($f$-test probability of $2\times 10^{-7}$). In summary, PL with large $N_{\rm H}=(1.1\pm0.2)\times 10^{21}\rm \ cm^{-2}$ or BPL with small $N_{\rm H}=(3.6\pm0.6)\times 10^{20}\rm \ cm^{-2}$ can fit the {\it XMM} spectra of J0131 well.

\subsection{Analysis of the NuSTAR data\label{sec:sec2_3}}
We next analyze the {\it NuSTAR} data. The {\it NuSTAR} spectra of J0131 are obtained with an $R$=$15''$ region. We use a small aperture in order to minimize contamination from S1. Background spectra are obtained using an $R=45''$ aperture in a source-free region on the same detector chip. The corresponding response files are generated with the {\tt nuproduct} tool.
We group the {\it NuSTAR} spectra to have at least 5 counts per bin and fit the spectra with a single PL model employing $l$ statistic \citep[][]{l92} in {\tt XSPEC}. The model describes the data well with $\Gamma=1.67\pm0.28$ and $F_{\rm 3-10\rm \ keV}=(7.1\pm1.4)\times 10^{-14}\rm \ erg\ cm^{-2}\ s^{-1}$ (Table~\ref{ta:ta1}).

The {\it NuSTAR} spectral analysis results for J0131 may sensitively depend on the extraction regions for the faint source and non-uniform background (Fig.~\ref{fig:fig1}), so we applied various source and background regions as a sanity check. We shift the source region by 1 pixel ($\approx2.5''$) in each direction to generate nine new regions and select nine background regions with varying size and location on the same detector chip. We then perform spectral analyses for the 81 combinations of the source and background regions, and find an average $\Gamma=1.64$ and the standard deviation 0.14. The former is similar to the best-fit value and the latter is significantly smaller than the fit uncertainty (Table~\ref{ta:ta1}). We therefore conclude that the {\it NuSTAR} fit results in Table~\ref{ta:ta1} represent the source spectra well.

The source is not detected by {\it NuSTAR} above $\sim$30\,keV, so we derive a flux upper limit in the 30--79\,keV band. Holding all spectral parameters fixed at the low-energy best-fit values (Table~\ref{ta:ta1}) except for the flux, we use the {\tt steppar} command in {\tt XSPEC} to find the 90\% flux upper limit $F_{30-79\rm \ keV}=4.8\times 10^{-13}\rm \ erg\ cm^{-2}\ s^{-1}$. This value is insensitive to the assumed spectral index.

\subsection{Joint fits of the broadband data\label{sec:sec2_4}}
To further constrain the broadband X-ray spectrum, we jointly fit all X-ray data including archival {\it Swift} exposures \citep[taken from][AR18 hereafter]{ar18}.
Note that we allow a {\tt const} in the model to account for cross normalization among the instruments; these cross-calibration constants are statistically consistent with 1 in the fits summarized below.
For the large $N_{\rm H}$ case, the combined data are well fit with a simple PL model
having spectral parameters consistent with the {\it XMM}-estimated values; a spectral break (i.e., BPL fit) is unnecessary ($f$-test $p$=0.4). For the small $N_{\rm H}$ case the fit is dominated by the high-count {\it XMM} data, with fit parameters consistent with those of a simple {\it XMM}. The results are summarized in Table~\ref{ta:ta1}, and the X-ray SEDs are presented in Figure~\ref{fig:fig3}:
BPL with small $N_{\rm H}$ (top) and simple PL with large $N_{\rm H}$ (bottom).

\section{Broadband SED characterization\label{sec:sec3}}
\begin{figure}
\centering
\includegraphics[width=85mm]{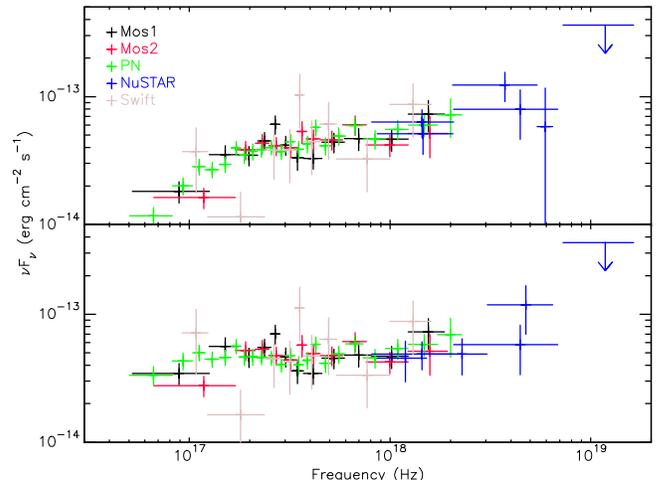} \\
\figcaption{0.2--30\,keV SEDs for the low-$N_{\rm H}$ BPL (top) and high-$N_{\rm H}$ PL (bottom) models.
\label{fig:fig3}}
\end{figure}

We supplement the X-ray data with archival radio-to-optical and {\it Fermi}-LAT data (\S~\ref{sec:sec3_1}).
The two SEDs differ only in the X-ray band. The low-$N_{\rm H}$ SED exhibits a rapidly rising trend
at low X-ray energies, while the high-$N_{\rm H}$ one shows a broad flat X-ray component (Fig.~\ref{fig:fig3}).
Our new SED measurements provide better coverage of the X-rays. However, the low-frequency ($<5\times10^{13}$\,Hz) and the gamma-ray SED are not well measured, limiting the accuracy of our inferred jet properties.

We assume that the broadband emission (Fig.~\ref{fig:fig4}) is produced by synchrotron and Compton emission by the jet electrons 
with an addition of disk emission in the optical band at $\sim10^{14}$\,Hz (\S~\ref{sec:sec3_2}).
The low-frequency $<10^{13}$\,Hz emission is attributed to synchrotron radiation, and
the X-rays ($10^{16}-10^{19}$\,Hz) and undetected gamma rays ($>10^{19}$\,Hz) are assumed to be produced by the SSC and disk EC processes, respectively (\S~\ref{sec:sec3_3} and \ref{sec:sec3_4}).
Alternatively, the X-ray emission may be attributed to the disk-EC process \citep[e.g.,][]{gtsg15}, and we consider this possibility as well (\S~\ref{sec:sec3_5}).

\subsection{Construction of the broadband SEDs of J0131\label{sec:sec3_1}}
\begin{figure*}
\centering
\includegraphics[height=175mm, width=130mm, angle=90]{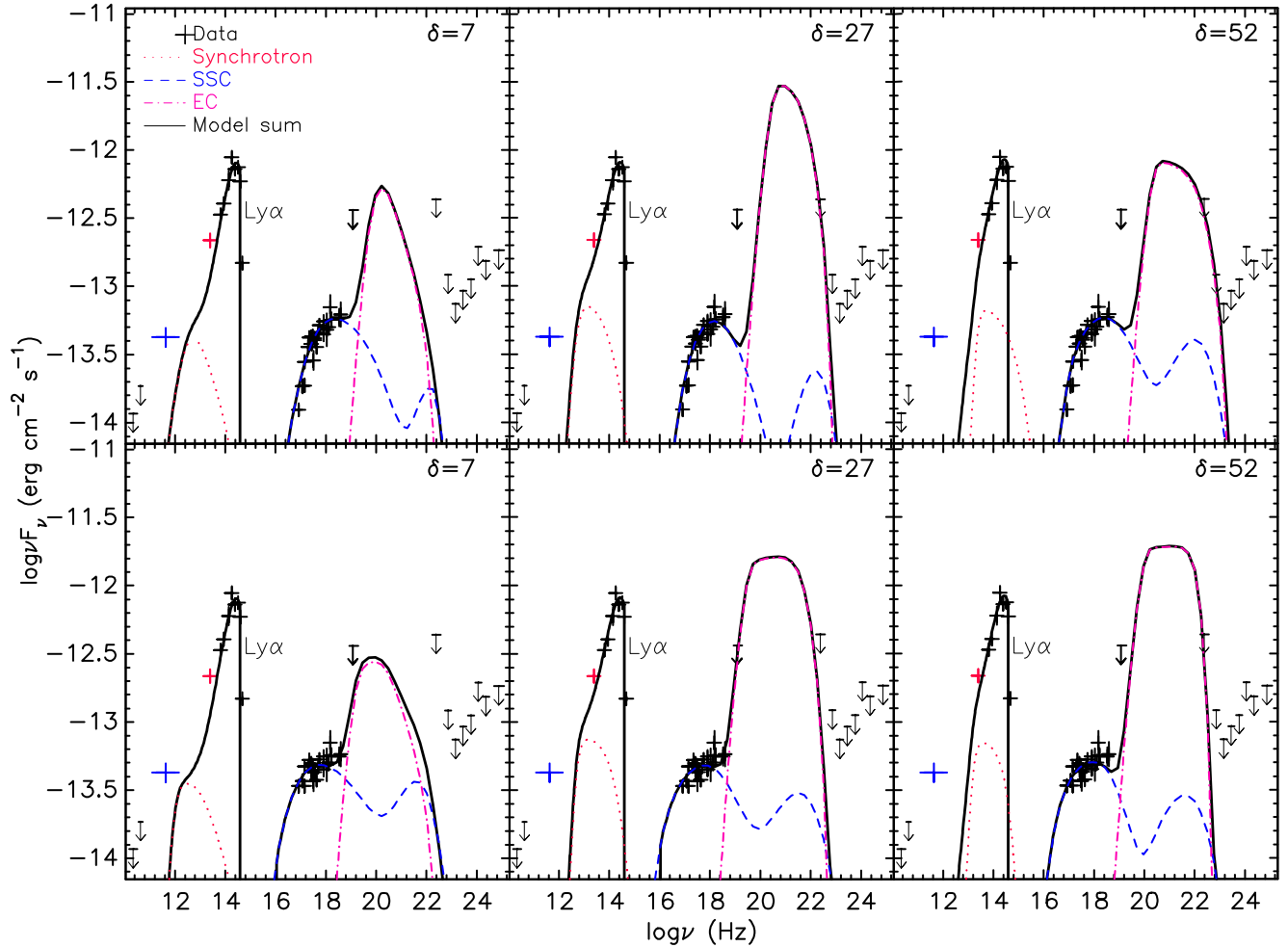}
\figcaption{Examples of broadband SED data and models ($\delta$=7, 27, and 52 from left) for J0131. The X-ray data are rebinned for better legibility and the {\it WISE}/W3 point is plotted in red.
{\it Top}: Low-$N_{\rm H}$ SED and models. {\it Bottom}: High-$N_{\rm H}$ SED and models. 95\% upper limits are plotted as down arrows. The SED models are displayed in black solid lines, and the model components are also presented: synchrotron (red dotted), SSC (blue dashed), and EC (pink dot-dashed) emission. The blue point is a {\it PLANCK} flux near the source, which can be taken as an upper limit.
\label{fig:fig4}}
\end{figure*}

The radio-to-optical measurements and {\it Fermi}-LAT upper limits are taken from previous works \citep[AR18;][]{zafg+17}. Thus, the principal SED updates are the revised X-ray measurements. Note that the X-ray spectra are produced by {\tt eeufspec} command in {\tt XSPEC} and corrected for the absorption. The resulting SEDs are shown in Figure~\ref{fig:fig4}.

The SED shape can provide important constraints on the jet properties such as the magnetic-field strength $B$ and $\delta$.
In particular, the peak frequencies of the SED humps are very useful for inferring $\delta$.
Without good SED coverage, the peak frequencies alone do not allow precise determination of $\delta$, but we can still use the frequency scaling presented in AR18 to roughly estimate the electron Lorentz factor ($\gamma_e$) and bulk Doppler factor: $\gamma_e \approx (\nu_{ssc,pk}^{obs}/\nu_{sy,pk}^{obs})^{1/2}$ and $\delta \approx (\nu_{EC,pk}^{obs}/\nu_{BB,pk}^{obs})^{1/2}/\gamma_e$, where $\nu_{X,pk}^{obs}$ is the ``observed'' peak frequency of the emission component ($X=ssc$ for synchro-self-Compton, $X=sy$ for synchrotron, $X=EC$ for external Compton, and $X=BB$ for disk emission). However, unlike QSO~J0906+6930 studied by AR18, the synchrotron SED of J0131 is poorly measured and $\nu_{sy,pk}^{obs}$ not well constrained. To make progress we assume that
$\nu_{sy,pk}^{obs}<10^{14}$\,Hz similar to other high-$z$ blazars, $\nu_{ssc,pk}^{obs}\approx 10^{18}$\,Hz, $\nu_{BB,pk}^{obs}\approx 3\times 10^{14}$\,Hz, and $\nu_{EC,pk}^{obs}\approx 10^{20-22}$\,Hz as an initial guess. We can then infer that $\gamma_e\ge100$ and $\delta \le 30$. These rough estimates, obtained based only on the observed peak frequencies, serve as a guide for SED model computation and are superseded by the more detailed modeling below.

\subsection{The SED model \label{sec:sec3_2}}
The model computation is performed with a blazar SED code adapted from \citet[][]{bms97}.
Electrons and positrons ($e^\pm$) with a power-law energy distribution
$dN_e/d\gamma_e\propto \gamma_e^{-p_1}$ are injected at an assumed height of $h=0.03$\,pc
from the black hole and cooled by radiation while they travel along the jet.
Hence, the energy distribution of the particles changes with time, and we follow the particle distribution and compute the emission for $10^7$\,sec.
In addition to the jet emission, the model computes the disk emission in the $10^{14-15}$\,Hz optical band assuming
a standard Shakura-Sunyaev disk \citep[][]{ss73} for $M_{\bullet}=1.5\times10^{10}M_\odot$ \citep[e.g.,][]{gtsg15,cgsc18},
where the disk luminosity is adjusted to match the optical SED; the disk emission is held fixed when modeling the jet emission below. The time-integrated model SED is then compared to the observations, and we adjust the model parameters to attain a match.
Note again that the X-ray SED is attributed to the SSC (\S~\ref{sec:sec3_3} and \ref{sec:sec3_4}) or EC (\S~\ref{sec:sec3_5}) process.

Because of the lack of SED measurements in some wavebands and covariance among the model parameters, they are not all well constrained; we thus also make the common assumptions of magnetic equipartition and $\Gamma_D=\delta$.
We further assume that the $e^\pm$ spectral index $p_1$ should be $\ge$1
since acceleration theory does not produce very hard energy distribution for the injected particles.

We start with previously-estimated parameters for J0131 (AR18) and the assumptions above, and vary the $e^\pm$ Lorentz factors $\gamma_e$'s, spectral index for the distribution $p_1$, particle number density $n_e$, the emitting volume $R_b$, and $\delta$ to match the observed SED. Because the parameter space is large and the parameters co-vary, simultaneously optimizing all parameters is unfeasible.
We therefore specify a $\delta$, Monte Carlo (MC) vary the other parameters, calculate the model $\chi^2$ in the IR (a {\it WISE}/W3 measurement at $\sim2.5\times10^{13}$\,Hz) and X-ray bands. In a downward descent, the next trial parameters are generated, by varying around values for the minimum $\chi^2$ model to that point. This procedure gives minimum $\chi^2$ model parameters for each $\delta$. Note that the optical points are disk-dominated and so are not used in this jet fit, except that the model must not exceed these optical fluxes.
We scan $\delta$ over the range 4--60 which is wide enough to cover the previous estimate of $\delta \approx $4--16 (AR18) and
the rough estimate presented above (\S~\ref{sec:sec3_1}).

Because the disk emission is strong compared to the synchrotron, the EC flux is large and often violates the LAT upper limits unless $B$ is also large so that the $e^\pm$ are efficient synchrotron/SSC radiators. Note that syncro-Compton cooling is efficient (large $B$ and $L_{\rm disk}$), so the synchrotron and SSC SEDs become fairly narrow. Examples of optimized models for specific $\delta$ are displayed in Figure~\ref{fig:fig4}, and the model parameters are presented in Table~\ref{ta:ta2}.

\begin{table*}[t]
\vspace{-0.0in}
\begin{center}
\caption{Parameters for the SED models in Figure~\ref{fig:fig4}}
\label{ta:ta2}
\vspace{-0.05in}
\scriptsize{
\begin{tabular}{lcccc|ccc} \hline\hline
Parameter       & Symbol        & \multicolumn{6}{c}{Value}     \\ \hline
Redshift & $z$ & \multicolumn{6}{c}{5.18}  \\
Black Hole mass ($M_\odot$) & $M_\bullet$ & \multicolumn{6}{c}{$1.5\times 10^{10}$} \\
Disk Luminosity (erg/s) & $L_{\rm disk}$ & \multicolumn{6}{c}{$1.1\times 10^{48}$}   \\ \hline
SED model &  & \multicolumn{3}{c}{low $N_{\rm H}$} & \multicolumn{3}{c}{high $N_{\rm H}$}  \\
Doppler factor  & $\delta$      & 7 & 27 & 52 & 7 & 27 & 52 \\
Magnetic field (G)      & $B$   & 15 & 30 & 110 & 19 & 41 & 72 \\
Comoving radius of blob (cm)    & $R'_b$& $7.1\times 10^{14}$ & $1.3\times 10^{14}$ & $2.2\times 10^{13}$ & $6.6\times10^{14}$ & $1.3\times10^{14}$ & $5.9\times10^{13}$ \\
Effective radius of blob (cm)   & $R'_E$$^{\rm a}$ & $4.8\times 10^{15}$ & $1.6\times 10^{15}$ & $4.7\times 10^{14}$ & $4.6\times 10^{15}$ & $1.6\times 10^{15}$ & $9.2\times 10^{14}$ \\
Electron density (cm$^{-3}$)    & $n_e$ & $2.7\times 10^{4}$ & $2.8\times10^5$ & $8.7\times 10^{6}$ & $9.4\times 10^{4}$ & $1.6\times 10^{6}$ & $6.0\times 10^{6}$ \\
Initial electron spectral index & $p_1$ & 2.7 & 2.9 & 3.0 & 2.6 & 2.9 & 3.0 \\
Initial min. electron Lorentz factor    & $\gamma_{\rm e, min}$ & $2.0\times10^{2}$ & $8.5\times10^1$ & $3.5\times10^1$ & $7.6\times10^1$ & $2.5\times10^1$ & $2.2\times10^1$ \\
Initial max. electron Lorentz factor    & $\gamma_{\rm e, max}$ & $2.4\times10^{3}$ & $9.2\times 10^2$ & $9.1\times 10^2$ & $2.6\times 10^3$ & $7.4\times 10^2$ & $3.5\times 10^2$ \\
Injected electron luminosity ($\rm erg\ s^{-1}$) & $L_{\rm inj}$$^{\rm b}$ & $1.3\times 10^{48}$ & $4.9\times10^{47}$ & $1.1\times10^{47}$ & $1.6\times10^{48}$ & $9.4\times10^{47}$ & $9.6\times10^{47}$ \\ \hline
Best fit $\chi^2$   & $\chi^2$ &  79  & 68 & 70  & 73 & 62 & 69 \\ \hline
\end{tabular}}
\end{center}
\vspace{-0.5 mm}
\footnotesize{
$^{\rm a}$ Effective radius of the elongated jet computed with $R'_E=(3{R'}_b^{2} t_{evol}c/4)^{1/3}$.\\
$^{\rm b}$ Jet power in the black hole rest frame.\\}
\end{table*}

\subsection{The low-$N_{\rm H}$ SED model \label{sec:sec3_3}}
As noted above, we optimize the SED models by scanning $\delta$ values and minimizing $\chi^2$ over the other parameters. We find a global minimum $\chi^2_{\rm min}=68$ (60 dofs) at $\delta=27$ (top panels of Fig.~\ref{fig:fig4}). The model $\chi^2$ converges better with low $\delta$'s than with high ones ($\ge$20); the high-$\delta$ models often fall into a local minimum. The general trend for the parameters is that $n_e$ (and $B$) increases while the other parameters decrease with increasing $\delta$. The parameters for a few $\delta$ values are presented in Table~\ref{ta:ta2}.

The behavior of the SED models for different $\delta$ values is complex but can be qualitatively explained as follows.
In the low-$\delta$ models, EC emission is relatively weak and the X-ray SED is easily matched by the SSC. So the $\chi^2$ is determined primarily by the {\it WISE}/W3 data point. Changing the model parameters (e.g., $B$) from the optimal values degrades the fit. Lowering $B$ requires larger $\gamma_e$'s (i.e., a shift of $dN_e/d\gamma_e$ in the high-$\gamma_e$ direction) to preserve the X-ray data match (i.e., $\nu_{ssc,pk}$), but $\nu_{sy,pk}$ lowers because $\nu_{ssc,pk}/ \nu_{sy,pk}\propto \gamma_e^2$, which worsens the match to the {\it WISE} point. On the other hand, increasing $B$ does the opposite, making the $\gamma_e$'s smaller and $\nu_{sy,pk}$ larger; so a better match to the {\it WISE} point is expected. However, $n_e$ also increases to preserve equipartition, which then lowers the synchrotron flux $F_{sy}$ for the given SSC flux $F_{ssc}$ (X-ray data) because $F_{ssc}/F_{sy}\propto n_e$. With EC this last relationship is not exact, but is adequate for the weak EC emission seen in low-$\delta$ high-$B$ models. This allows a reliable determination of a 90\% lower limit $\delta>7$ ($\Delta \chi^2=11$ for 6 dofs).

As $\delta$ increases, the separation between the SSC and EC peaks grows. With the SSC component set by the X-ray measurements, EC emission encroaches on the LAT band in high-$\delta$ models (Fig.~\ref{fig:fig4}). As noted above, lowering $B$ increases $\gamma_e$'s, pushing the EC emission to higher flux and frequencies which degrades the fit. Increasing $B$ typically improves the fit by lowering the EC flux and frequency. Although the {\it WISE} point remains a concern in high-$B$ models, unlike the low-$\delta$ case these models can accommodate it because we no longer have $F_{ssc}/F_{sy}\propto n_e$ (EC emission dominates). Thus we find an acceptable fit even for the highest $\delta$=60 investigated.

However, continued $B$ (and $\delta$) increase will create model problems; (1) $\nu_{sy, pk}$ becomes too high to accommodate the {\it WISE} point, (2) when the EC emission is sufficiently suppressed, $F_{ssc}/F_{sy}\propto n_e$, and we must lower the synchrotron emission to keep the X-ray (SSC) data match. This latter situation only applies at very high $\delta>60$. If we assume that J0131 is a low-synchrotron-peaked (LSP) source and has $\nu_{sy,pk}<10^{14}$\,Hz, $\delta\le30$ may be inferred.

\subsection{Modeling the high-$N_{\rm H}$ SED \label{sec:sec3_4}}
Because of rapid cooling, the SSC emission is narrow and the high-$N_{\rm H}$ SED (Fig.~\ref{fig:fig3}), which is flat across the 0.2--30\,keV band, is hard to match with SSC only. However, no X-ray break at a low energy is needed (Fig.~\ref{fig:fig3} bottom) and the minimum $e^\pm$ Lorentz factor $\gamma_{\rm e, min}$ can be small. This moves the EC component to lower frequencies compared to low-$N_{\rm H}$ models (\S~\ref{sec:sec3_3}), and the high {\it NuSTAR} points above 10\,keV can be accommodated by the rising part of the EC component (Fig.~\ref{fig:fig4} bottom). Therefore, for modest $\delta$'s both the {\it WISE}/W3 point and the {\it NuSTAR} data are well matched, and the minimum $\chi^2=62$ (for $\delta$=28) is substantially lower than that of the best low-$N_{\rm H}$ SED fit (Table~\ref{ta:ta2}). Clearly a deeper {\it NuSTAR} exposure with a detection above 30\,keV would be a good test of these models. The general parameter trends are similar to the low-$N_{\rm H}$ case; $n_e$ increases while the other parameters decrease with increasing $\delta$.

Comparing across $\delta$ in the high-$N_{\rm H}$ case gives results similar to low-$N_{\rm H}$; low-$\delta$ models conflict with the {\it WISE}/W3 point, and high-$\delta$ values under-predict the {\it NuSTAR} flux, although the hard X-ray {\it NuSTAR} points are better matched with the high-$N_{\rm H}$ SED, as noted above. As for the high-$N_{\rm H}$ SED models, our strongest conclusion is a lower limit of $\delta>6$. Large $\delta$ values produce only modest $\chi^2$ increase (e.g., at $\delta=50$ the increase is only $\Delta \chi^2\le$9 for 6 dofs; 68\%), so these large values, while sub-optimal, are still acceptable.

Note that the parameters presented in Table~\ref{ta:ta2} are not unusual for blazar jets, but the inferred $B$'s are large compared to $\le$a few Gauss often used in leptonic models for blazars. It is hard to estimate $B$ in blazar jets model-independently (i.e., from fundamental physics), but $>$10\,G has been invoked in leptonic models \citep[e.g.,][]{dclb+14} and even higher $\sim$100\,G values are used in lepto-hadronic models \citep[e.g.,][]{bbpc16}.

\subsection{EC models for the SED \label{sec:sec3_5}}
We assumed above (\S~\ref{sec:sec3_3}) that the X-ray emission is produced by the SSC process, as is usually inferred for blazars. However, in principle disk-EC might dominate, so we check this possibility.

\begin{figure}
\centering
\includegraphics[width=58mm, angle=90]{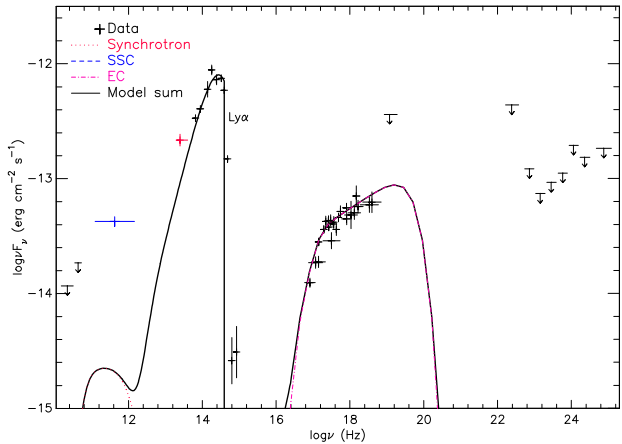} 
\figcaption{The broadband SED for the low-$N_{\rm H}$ case of J0131 (data points) and an EC model ($\delta=2.2$) in which X-ray emission is assumed to be produced by the EC process. The SSC emission is too low to be shown in this plot.
\label{fig:fig5}}
\end{figure}

Here suppressing SSC emission with low $B$, $n_e$ and/or large $R_b$ also suppresses the synchrotron flux. An example (very low $\delta=2.2$) of such models is presented in Figure~\ref{fig:fig5} along with the low-$N_{\rm H}$ SED. It is easy to match the X-ray SED shape by the EC emission, but the synchrotron emission is very weak and so the {\it WISE}/W3 point is brighter than the pure disk flux in these models. Increasing $\delta$ will make the synchrotron emission even weaker compared to the EC (X-ray) and does not improve the fit. These fits are inferior (e.g., $\chi_{\rm min}^2\approx78$ regardless of $\delta$) to the SSC-dominate models in \S~\ref{sec:sec3_3} and \ref{sec:sec3_4} unless the W3 point is attributed to
another emission component (e.g., a dust torus).
A similar conclusion can be drawn for the high-$N_{\rm H}$ SED. We therefore do not consider these models further.

We note, however, that EC-dominated one-zone models of \citet{gtsg15} with an additional blackbody component for the W3
point appear to explain the SEDs. If it can be shown that the W3 data are produced by
blackbody emission, more detailed EC modeling needs to be carried out; additional $10^{13-14}$\,Hz IR data would be very helpful.

\section{Discussion and Conclusion\label{sec:conclusion}}

We investigated the X-ray emission of the high-$z$ blazar J0131 using new {\it NuSTAR} and {\it XMM} observations, improving on previous studies which relied on limited {\it Swift} data. We found that the X-ray spectrum can be described by a simple $\Gamma=1.9$ PL with an absorbing column $N_{\rm H}=1.1\times 10^{21}\rm \ cm^{-2}$ or by a less-absorbed ($N_{\rm H}=3.6\times 10^{20}\rm \ cm^{-2}$) BPL with $\Gamma_{\rm l}=0.5$, $\Gamma_{\rm h}=1.8$, and $E_{\rm b}=0.7$\,keV. Intriguingly the hard X-ray ($>$10\,keV) data seem to suggest that the spectrum is not falling in that band, unlike the high-$z$ blazars B2~1023+25 and QSO~J0906+6930 (AR18), although with current data the rise is not highly significant.

We constructed broadband SEDs of the source for both the low-$N_{\rm H}$ and high-$N_{\rm H}$ X-ray cases by supplementing the new X-ray measurements with archival data, and model the SEDs using a synchro-Compton scenario.
Because J0131 is not detected by the {\it Fermi} LAT and there is no sensitive observation in the lower-energy gamma-ray band,
our {\it NuSTAR} measurements provide important constraints on the Compton peak.
The X-ray flux could be modeled via two different scenarios: SSC or disk EC. While the EC models can describe the X-ray emission, the models have very low synchrotron flux and do not explain the {\it WISE}/W3 point. We considered this less likely and so have focused on the SSC case.

We used a MC technique to explore values of the bulk Doppler factor in these SSC models, finding $\delta>$7 and $\delta>$6 (90\%) for the low- and high-$N_{\rm H}$ SEDs, respectively. These lower bounds for the J0131 jet seem reasonable given that radio observations of other high-$z$ blazars infer modest bulk Doppler factors for their radio jets \citep[e.g., $\delta_{\rm radio}=13$ and 6 for B2~1023+25 and QSO~J0906+6930;][]{fpfg15,amzf+20}. However, our lower bound is quite sensitive to the single {\it WISE}/W3 data point and would change if the IR flux were lower or higher during the X-ray observation. 
Interestingly, the disk emission model also under-predicts the {\it WISE}/W2 data point at $\approx 7\times 10^{13}$\,Hz (Fig.~\ref{fig:fig4}). This is also seen in a Kerr black hole disk emission model for J0131 \citep[Fig.~8 of][]{cgsc18}. If there is a synchrotron contribution at this frequency, this would tend to increase our $\delta$ lower bound. However, since the spectral shape of disk emission around super-massive black holes is poorly known we do not attempt to match the W2 data point with our SED model. Longer wavelength observations will provide more reliable constraints on the synchrotron SED peak.

Some uncertainties are also due to modeling assumptions. For example, if the emission region lies father away from the black hole than our assumed 0.03\,pc, the disk EC emission is suppressed and even higher $\delta$ models might be allowed. However, other external emission (dusty torus and broad line region BLR) may contribute to the EC emission. This is only poorly bounded by the LAT upper limits. The distance to the emission region is also related to the large $B$ which we inferred from the models (Table~\ref{ta:ta2}). These models demand large $B$ to suppress the EC emission, but a smaller $B$ would be allowed if the region were located farther away. Also, the black hole mass determining the disk temperature and luminosity is also uncertain, although mass and disk SED we used are similar to those inferred in a J0131 Kerr hole disk model \citep[$1.23\times 10^{10}M_\odot$;][]{cgsc18}. In general with smaller hole mass, the temperature and luminosity of the disk grow, making the EC component stronger; this is a weak $T\propto M^{-1/4}$ effect. Nevertheless, our $\delta>6$ lower bound should not be altered since EC emission contributes little to the low-$\delta$ models. 
Note that the parameters in Table~\ref{ta:ta2} are not very extreme. But given that $B$ and $\delta$
for other high-$z$ ($z$=3--4) blazars are inferred
to be modest \citep[$B$=1--2\,G and $\delta=$10--16;][]{pacg+20},
the low-$\delta$ (e.g., $<$20) models may be preferred for J0131.

Using the $\delta$ constraints and the formula given in AR18, we infer that there should be $>$100 (low $N_{\rm H}$) and $>$70 (high $N_{\rm H}$) sources similar to J0131 (see AR18 for details). We cannot yet obtain a firm upper bound on $\delta$ directly from the modeling, but assuming that J0131 is an LSP we could set a weak bound $\delta\le30$.
Then, the inferred number of similar (i.e., J0131-like) blazars
would be 230--420 (low $N_{\rm H}$) or 170--360 (high $N_{\rm H}$).
This may not be a large improvement compared to the previous results (AR18). However, in the previous work
a formal SED fit was not attempted for J0131 due to the lack of X-ray data points,
and the model was matched to the SED only by eye. Hence, the new constraints
on $\delta$ based on formal fits, made possible by the high-quality X-ray data,
provide a more accurate population estimate.

Of course, our parameter constraints and population numbers can be tightened by a more complete SED. Sub-mm/IR measurements can pin down the synchrotron flux. The X-ray is the key to the SSC component. A better understanding of the absorption might be obtained with more sensitive soft X-ray or even optical absorption studies. This indirectly affects the inferred X-ray power law, which compared to the synchrotron peak gives us our $\delta$ bounds; high sub-mm flux implies larger $\delta$. Since the EC peaks of high-$z$ blazars are expected to lie below the LAT band, the best measurement of the EC component can be made by extending our hard X-ray/soft gamma-ray measurements; a $>10$\,keV break to a harder spectrum implies an EC peak at low frequency (and a small $\delta$). Hence, future ALMA, {\it JWST}, {\it NuSTAR}, and {\it AMEGO} \citep[][]{amego19} observations are warranted to help pin down J0131's beaming and let this member of the very small $z>5$ jet-dominated QSO sample  constrain the high-$z$ blazar population.

\acknowledgments
We thank the anonymous referee for the careful reading of the paper and
insightful comments.
This research was supported by Basic Science Research Program through
the National Research Foundation of Korea (NRF)
funded by the Ministry of Science, ICT \& Future Planning (NRF-2017R1C1B2004566).

\bigskip
\vspace{5mm}
\facilities{{\it XMM}, {\it NuSTAR}}
\software{HEAsoft (v6.26.1; HEASARC 2014),
          XMM-SAS \citep[v20180620; ][]{xmmsas17},
          XSPEC \citep[][]{a96}
          }

\bigskip
\bibliographystyle{apj}
%\expandafter\ifx\csname natexlab\endcsname\relax\def\natexlab#1{#1}\fi
%  \input{J0131.bbl}
\bibliography{GBINARY,BLLacs,PSRBINARY,PWN,STATISTICS,FERMIBASE,COMPUTING,INSTRUMENT,ABSORB}
\end{document}